\documentclass[pra,showpacs,floatfix,amsmath,amsfonts, twocolumn]{revtex4}
\usepackage{graphicx}
\usepackage[utf8]{inputenc}
\usepackage{dcolumn}
\usepackage{bm}
\usepackage{amsmath}
\usepackage{amssymb}
\usepackage{wasysym}
\usepackage{color}

\begin{document}

\title{Dynamics of vortex-antivortex pairs and rarefaction pulses in liquid light}
\author{David Feijoo, Angel Paredes and Humberto Michinel}
\affiliation{Departamento de F\'\i sica Aplicada,
Universidade de Vigo, As Lagoas s/n, Ourense, ES-32004 Spain\\
}

\begin{abstract}
We present a numerical study of the cubic-quintic nonlinear Schr\"odinger 
equation in two transverse dimensions, relevant for the propagation of light
in certain exotic media. A well known  feature of 
the model is the existence of flat-top bright solitons of fixed intensity, whose dynamics
resembles the physics of a liquid. They support traveling wave solutions, consisting
of rarefaction pulses and vortex-antivortex pairs. 
In this work, we demonstrate how the vortex-antivortex pairs can be generated in bright
soliton collisions displaying destructive interference followed by a snake instability.
We then discuss the collisional dynamics of the dark  excitations
for different initial conditions. We describe a number of distinct phenomena
including vortex exchange modes, quasi-elastic flyby scattering, 
soliton-like crossing, fully inelastic collisions and rarefaction pulse merging.

\end{abstract}


\maketitle


\section{Introduction}

The synergy between competing nonlinearities
in the Schr\"odinger equation can give rise to very interesting dynamics
\cite{PhysRevLett.102.203903,setzpfandt2009competing}, including, for instance,  solitons
 \cite{PhysRevA.83.053838,Laudyn:15} and phase transitions 
 \cite{PhysRevLett.116.163902,0295-5075-98-4-44003}.
In this paper, we provide novel insights on the (focusing) cubic- (defocusing) quintic model, which has 
been thoroughly studied in the context of nonlinear optics 
\cite{mihalache1988exact,pushkarov1996bright,dimitrevski1998analysis}, 
where it was shown that large power solitons have neat similarities with regular liquids,
thereby motivating the term ``liquid light'' \cite{michinel2002liquid}.
The same equation
has been applied
in other frameworks too, see {\it e.g},
\cite{josserand1997coalescence,muryshev2002dynamics,khaykovich2006deviation,carretero2008nonlinear,davydova2003two}.

The cubic-quintic equation is an appropriate model for the propagation of 
light in certain optical materials, 
see for instance \cite{smektala2000non} and references in \cite{caplan2012existence}.
It has also been used as an approximation to the process of filamentation
\cite{piekara1974analysis,centurion2005dynamics,novoa2010filamentation}.
Recent experimental advances reinforce the significance of new theoretical studies.
Despite damping, (limited) soliton propagation has been observed in carbon disulfide
\cite{falcao2013robust}. Furthermore, the droplet-like behavior of cubic-quintic propagation
has been demonstrated in atomic gases at
 low optical powers \cite{wu2013cubic,wu2015solitons},
using quantum coherence and interference as proposed in
 \cite{michinel2006turning,alexandrescu2009liquidlike}.
Other setups in which the fifth order nonlinearity can be enhanced through quantum
effects comprise Rydberg atoms \cite{bai2016enhanced} and quantum dots
\cite{peng2014tunneling,tian2015giant}.
Confinement and guiding of light in a (defocusing) cubic- (focusing) quintic
has also been reported \cite{reyna2016guiding}.

In the cubic-quintic model, there is a one-parameter family  of
form-preserving 
 traveling dark wave solutions within a critical bright background, 
 which was computed in 
 \cite{paredes2014coherent} following the numerical methods of \cite{chiron2016travelling}.
For small velocities, it consists of vortex-antivortex pairs of charges $\pm 1$ (we will make
a usual abuse of language and refer to ``velocity'' for what in the optical setup corresponds
to the propagation angle with respect to the axis).
For larger subsonic velocities, the solutions are
rarefaction pulses, namely dark blobs without vorticity. 
The fainter the pulse is, the faster it moves within the bright background.
This family of solutions is similar to the one existing for third order
defocusing nonlinearity
 \cite{jones1982motions,jones1986motions,berloff2004motions,bethuel2009travelling}.
 Rarefaction pulses
 should not be confused with the unstable quiescent bubbles of
  \cite{barashenkov1988soliton,barashenkov1989stability}.

A separate issue is how these dark soliton-like excitations can be generated dynamically.
In the context of Bose-Einstein condensates (BEC), they have been generated by phase imprinting
\cite{proud2016jones}. In the framework of superfluids, it was shown that they can appear
when the fluid flows past an obstacle \cite{josserand1995cavitation}, a process that in optics
can be mimicked by the nonlinear interaction with an incoherently coupled beam 
\cite{feijoo2014drag} and in BECs with a laser beam (see \cite{mironov2010structure} 
and references therein). 

A remarkable result of \cite{paredes2014coherent} is that, for the liquid of light, rarefaction
pulses can 
be generated by interference in the collision of two bright solitons of very different sizes and powers.
The analogy with bubbles in fluids motivates the usage of the term cavitation for this kind of 
process.
The produced caviton excitation propagates within the large  soliton and can exit it 
becoming a bright soliton again. This bright-dark-bright conversion is familiar in one dimension,
see {\it e.g.} \cite{kim2000soliton,garralon2013numerical}, but it is a distinctive 
feature of the cubic-quintic equation
in two dimensions. This peculiarity facilitates the creation of dark traveling waves in a controlled manner
from initial conditions comprising only bright solitons.
With three initial bright solitons, two separate traveling waves can be created within the same
fluid. 

The natural question that we address in the
present paper is how these traveling waves interact with each other.
 It would be really interesting to 
implement this kind of processes in experimental setups as those described in 
\cite{falcao2013robust,wu2013cubic,wu2015solitons}. 
For the case of defocusing cubic nonlinearity, the dynamics of the dark excitations
in a nontrivial background was analyzed in 
\cite{smirnov2012dynamics,mironov2012propagation,mironov2013scattering}
and their interaction with a single vortex in
\cite{smirnov2015scattering}.

In section II, we fix notation and review some features of the cubic-quintic model.
 In section III, we show
that vortex-antivortex pairs can be produced by a soliton-soliton collision. 
Sections IV-VI describe the result of our simulations concerning dark wave interactions.
We discuss in turn the collision of two vortex-antivortex pairs, that of a rarefaction pulse
with a vortex-antivortex and that of two rarefaction pulses.
In section VII we outline our conclusions and make some final remarks.
The supplemental material \cite{suppl} contains animations for all of the examples of dynamical evolution
that are presented along the paper and a few extra illustrative cases.

\section{Solitons and traveling waves}

In this section we briefly review well-known results concerning the cubic-quintic
model in order to provide the basic ingredients for the following.
In the paraxial approximation, the canonical
equation 
governing the wave amplitude $\psi(x,y,z)$ reads:
\begin{equation}
i\partial_z \psi = - (\partial_x^2 + \partial_y^2)\psi - (|\psi|^2 - |\psi|^4) \psi .
\label{CQeq}
\end{equation}
A refractive index of the form $n=n_0+n_2 I + n_4 I^2$ has been assumed, where $I \sim |\psi|^2$ is the
intensity. It is straightforward to check that the equation in terms
of physical quantities can be rescaled to the dimensionless variables of Eq. (\ref{CQeq})
 without loss of generality
as long as $n_2>0$, $n_4<0$.

There are stable solitary waves 
of the form $\psi= e^{i\beta z}f(r)$ with $\lim_{r\to \infty} f(r)=0$
which we laxly
call bright solitons, as it is customary in the literature. 
The numerical study of \cite{dimitrevski1998analysis,michinel2002liquid}
shows that there are solutions for $0<\beta < \beta_{cr} = \frac{3}{16}$.
The power $P=2\pi \int r f(r)^2 dr$ grows monotonically with $\beta$ in the range
$P_0 < P < \infty$ where $P_0$ is the minimal value that leads to self-trapping.
For small $\beta$, the function $f(r)$ is bell-shaped. Near the
$\beta_{cr}$
 eigenvalue cutoff
\cite{prytula2008eigenvalue}, $f(r)$ tends to a flat-top profile.
This means that $f \approx \Psi_{cr} = \frac{\sqrt3}{2}$ for $r<r_{sol}$ and
around the soliton radius $r_{sol}$ there is a quick drop to
$f \approx 0$ for $r>r_{sol}$. This limit is the liquid-like phase, in which the 
soliton resembles a fluid with constant density and fixed surface tension
subject to the Young-Laplace equation \cite{novoa2009pressure}.

\begin{figure}[h!]
\begin{center}
\includegraphics[width=\columnwidth]{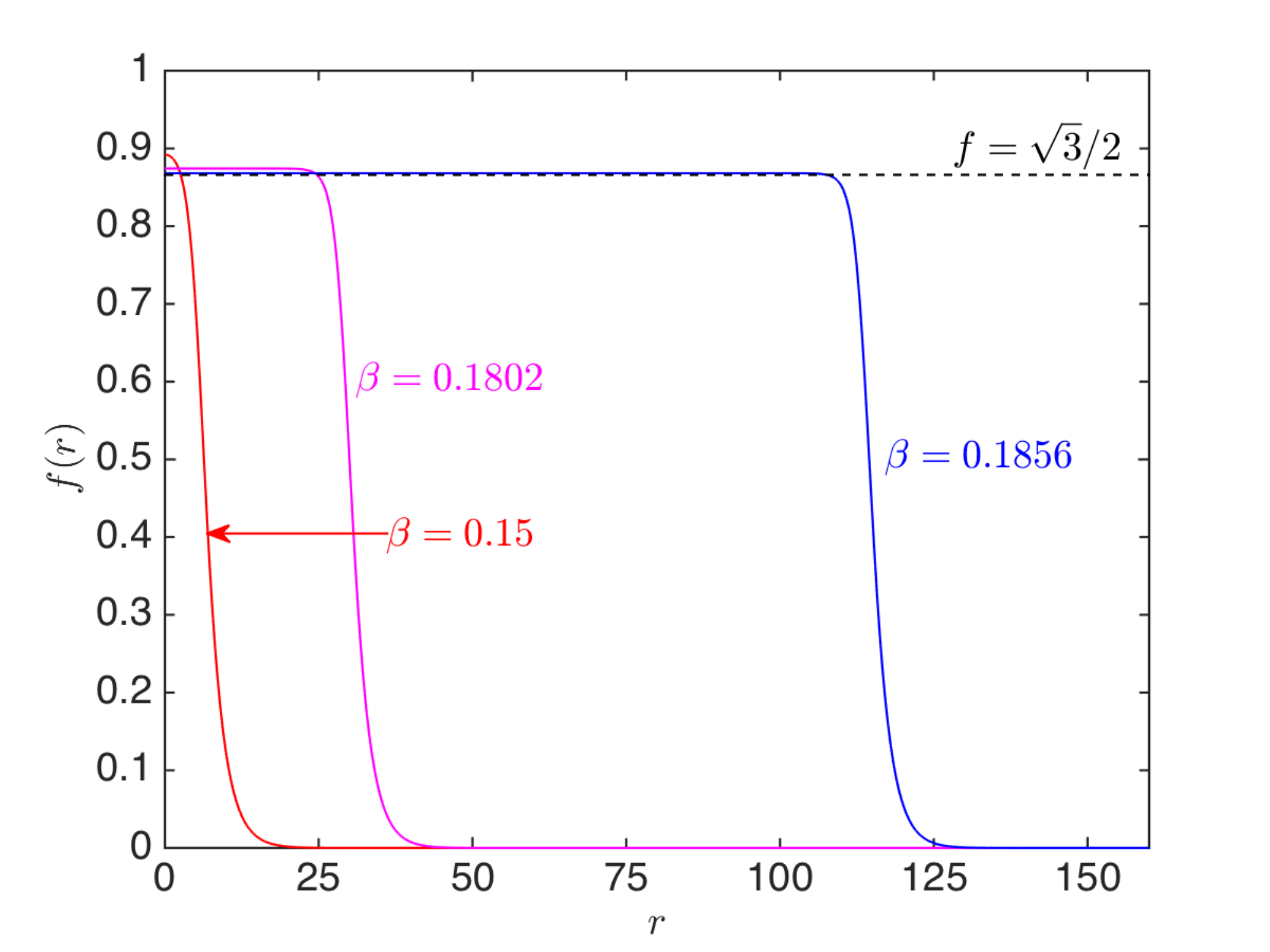}
\end{center}
\caption{(Color online) Radial profiles of the three solitons used to define the initial conditions in
the examples below. Their powers $P=2\pi \int_0^\infty r f(r)^2 dr$ are 
$P|_{\beta=0.15}=86.0$, $P|_{\beta=0.1802}=2055.5$, $P|_{\beta=0.1856}=30620$.}
\label{fig1}
\end{figure}

We will use these bright solitons to define the initial conditions of 
simulations in the following sections, by  considering:
\begin{eqnarray}
\psi|_{z=0}&=& f_1(|{\bf x}|) + f_2(|{\bf x}-{\bf x_2}|)\exp\left(i\frac{{\bf v_2}\cdot {\bf x}}{2}
+ i \phi_2\right)
+ \nonumber\\
&+& f_3(|{\bf x}-{\bf x_3}|)\exp\left(i\frac{{\bf v_3}\cdot {\bf x}}{2}
+ i \phi_3\right)
\label{init}
\end{eqnarray}
where the ${\bf x_i}$ are the initial positions of the solitons,
${\bf v_i}$ their initial velocities and $\phi_i$ their initial phases.
Boldface symbols are two-dimensional vectors.
The $f_i(.)$ are the soliton profiles, where $f_1$ is a flat-top soliton,
corresponding to the liquid where the dynamics  takes place
and $f_2$, $f_3$ are smaller solitons that  dynamically generate the dark excitations.
In Fig. \ref{fig1}, we plot the profiles of the particular solitons that will be used in
all the examples below.

Let us now turn to the dark traveling waves \cite{paredes2014coherent}.
They are form-preserving solutions of Eq. (\ref{CQeq}) moving at constant speed $U$
in, say,  the $x$-direction, embedded in an infinite liquid. 
Inserting the ansatz $ \psi(x,y,z)=e^{i\,\beta_{cr}\,z}\Psi(\eta,y) $ \cite{jones1982motions}.
where $\eta=x-Uz$, we can write:
\begin{equation}
 iU\partial_\eta \Psi= (\partial_\eta^2 + \partial_y^2)\Psi + 
 \left(|\Psi|^2 - |\Psi|^4-\frac{3}{16}\right)\Psi
 \label{EQtravel}
 \end{equation}
subject to the boundary condition
$
\lim_{\eta^2 + y^2 \to \infty}\Psi = \Psi_{cr}=\frac{\sqrt3}{2}$.
There is a family of solutions parameterized by $0<U<\frac{\sqrt3}{2}$.
For small $U$ they are vortex-antivortex pairs, with 
$|\psi|^2=0$ at the phase singularities. When $U$ grows, the vortex and antivortex
merge into a rarefaction pulse, whose $|\psi|^2$ is nowhere vanishing.
It is important to remark that the transition  is completely smooth and, roughly, one can think of the
rarefaction pulse as a bound state of vortex and antivortex. In fact, under nontrivial 
dynamical evolution both kinds of eigenstates can transform into each 
other \cite{mironov2012propagation,smirnov2015scattering}.

\begin{figure}[h!]
\begin{center}
\includegraphics[width=\columnwidth]{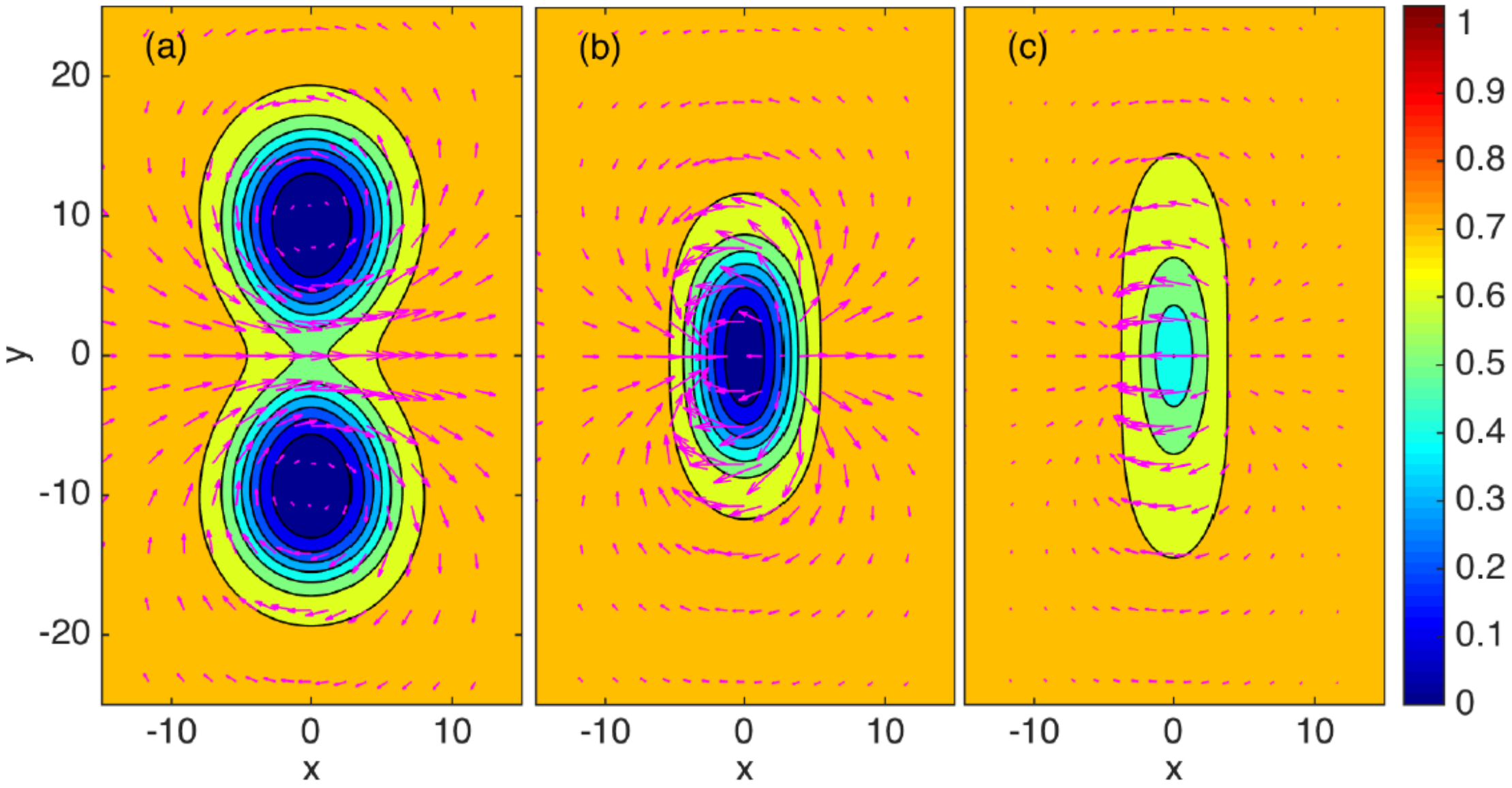}
\end{center}
\caption{(Color online) Three numerically computed traveling waves with $U=0.11$
(vortex-antivortex pair, panel (a)), $U=0.35$ (rarefaction pulse, panel (b)) and
$U=0.71$ (faint rarefaction pulse, panel (c)).
The dark excitations are moving rightwards.
The color scale displays
the density $|\psi|^2$ and the arrows represent the current density ${\bf j}$.}
\label{fig2}
\end{figure}

An interesting quantity is the current density which, in the hydrodynamical picture,
represents the flow of the fluid.
\begin{equation}
{\bf j}= \frac{1}{i} (\psi^* {\bf \nabla} \psi - \psi {\bf \nabla} \psi^*)  
\label{flow}
\end{equation}
The ${\bf j}$ is essential to understand how the dark excitation modifies the
medium around it and therefore to understand
the interaction between traveling waves. In figure \ref{fig2}, we depict
this quantity for three examples of traveling waves.

Momentum and energy are conserved quantities defined by:
\begin{eqnarray}
p &=& \frac{1}{2i}\int \left[(\Psi^* - \Psi_{cr})\partial_x \Psi
-(\Psi - \Psi_{cr})\partial_x \Psi^* \right] dxdy
\nonumber
\\
E&=& \int |{\bf \nabla} \Psi|^2 dxdy +
\frac13
\int |\Psi|^2\left(|\Psi|^2-\Psi_{cr}^2   \right)^2 dxdy
\label{pE2}
\end{eqnarray}
Within the family of solutions, one can check that $U=\partial E/\partial p$
and three virial identities are satisfied \cite{jones1982motions,jones1986motions,paredes2014coherent}.

The analyisis in the coming sections results from the numerical integration of
Eq. (\ref{CQeq}) with initial conditions (\ref{init}). The computations are done
using a standard split-step beam propagation method
\cite{agrawal2007nonlinear}. 
The evolution associated to the
non-derivative terms is computed with a fourth order
Runge-Kutta method. The plotted figures are built using grids of 800$\times$600
points. We have checked convergence of the method by comparing results with different
 grids in $(x,y)$ and steps in $z$.

\section{Coherent generation of vortex-antivortex pairs}

In \cite{paredes2014coherent}, it was shown that a rarefaction pulse
can appear when two coherent bright solitons meet with appropriate relative
velocity and phase. 
 Roughly speaking, destructive interference generates a void at the collision point
 which can acquire the necessary velocity thanks to the incoming momentum.
 Although, definitely, an exact solution of
(\ref{EQtravel}) is not realized in the dynamical process,
the resulting robust dark excitation  can indeed be identified with
 a traveling wave solution. This fact was checked in  \cite{paredes2014coherent} by 
comparing the dispersion relations. Even if the size of the medium
(the large soliton) is not infinite, it can support the traveling wave if it
is much larger than the dark structure.

\begin{figure}[h!]
\begin{center}
\includegraphics[width=\columnwidth]{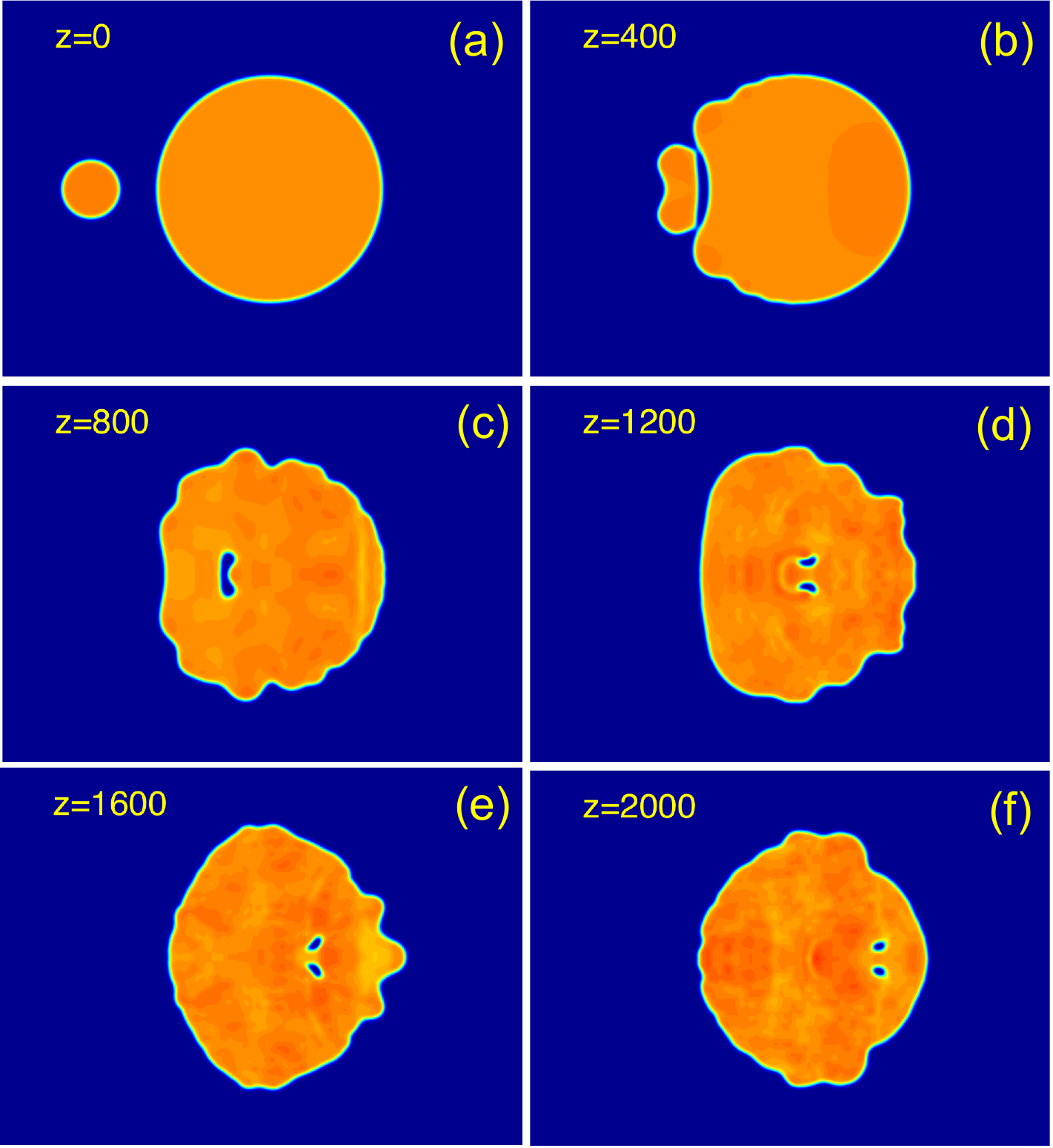}
\end{center}
\caption{(Color online) The encounter of two bright solitons giving rise to a traveling vortex-antivortex
pair. Initial conditions have ${\bf x_2}=(-180,0)$, ${\bf v_2}=(0.2,0)$, $\phi_2=5$.
The large soliton is the one with $\beta_1=0.1856$ and the smaller one has $\beta_2=0.1802$.
The color code for $|\psi|^2$ is as in Fig. \ref{fig2} and
the range of the axes is $x \in [-270,270]$, $y \in [-190,190]$. 
}
\label{fig3}
\end{figure}

In this section, we show that a similar process can result in the formation of 
a vortex-antivortex pair. In fact, the difference with \cite{paredes2014coherent}
is simply that the incomnig soliton has to be larger. What happens is that
during a collision in phase opposition, an elongated dark region is created. It cannot be stable
because there are no rarefaction pulse solutions of similar size. Consequently, it
evolves and decays through a snake instability giving rise to the separate vortex
and antivortex, which move forward together with a given velocity $U$.
Since the resulting configuration is not exactly equal to the stationary solution,
the dark regions can change, reconnect and split again. However
the vortex-antivortex profile becomes apparent after long enough propagation in $z$.
An example is depicted in figure \ref{fig3}.
Obviously, the third soliton of (\ref{init}) is not included in the initial condition.

\begin{figure}[h!]
\begin{center}
\includegraphics[width=\columnwidth]{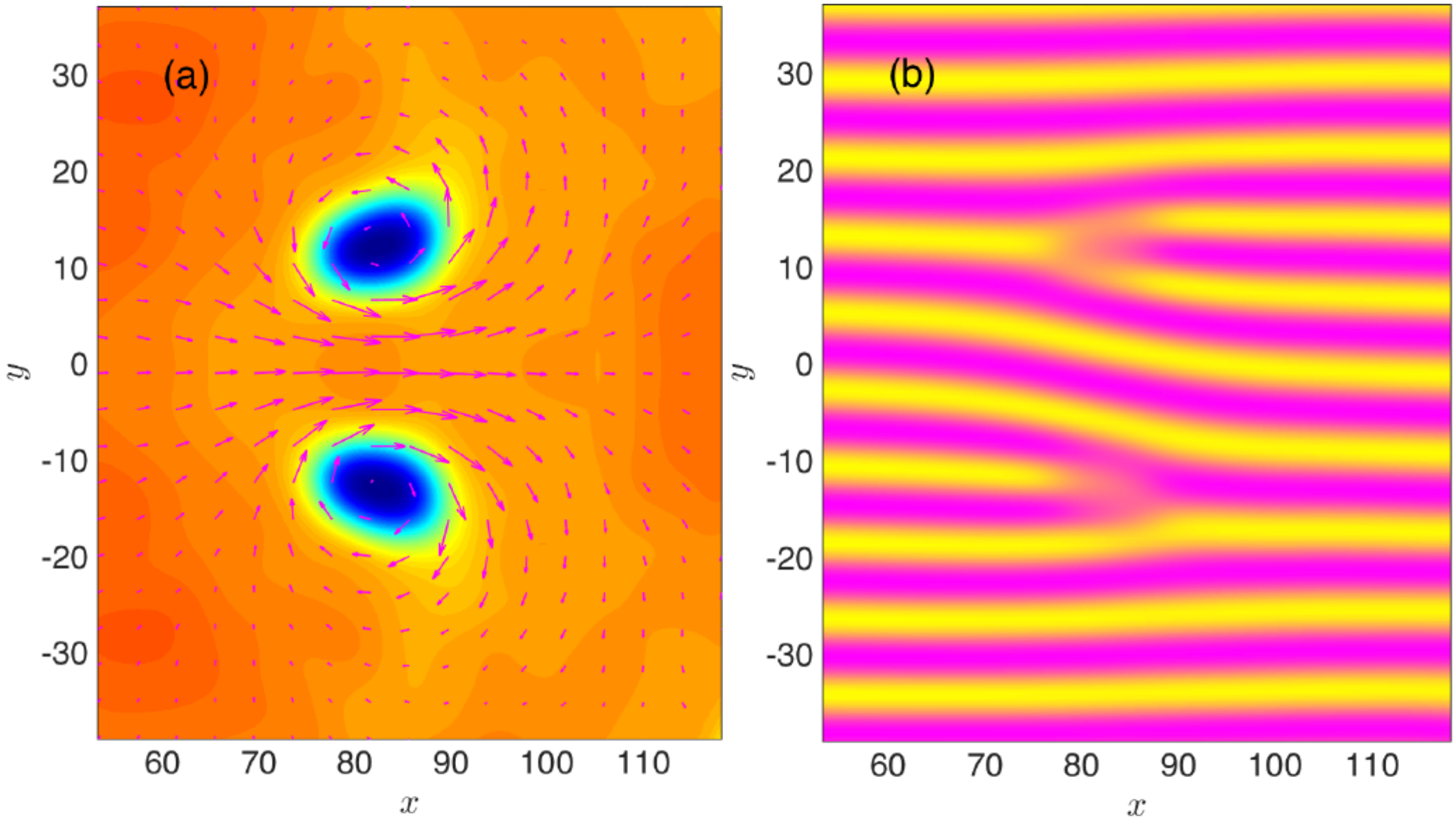}
\end{center}
\caption{(Color online) Phase structure of the wavefunction. The plot corresponds to $z=2000$,
panel (f) of
figure \ref{fig3}. The region of the vortex-antivortex has been enlarged .
 Panel (a) depicts $|\psi|^2$ with the same color scale of Fig. \ref{fig2} 
 and the  arrows are a quiver diagram for ${\bf j}$ showing flows similar
to figure \ref{fig2}. Panel (b) 
corresponds to the interference pattern 
 with a plane wave: $|\psi(x,y,z=2000)+7 \exp{(-100 i y)}|^2$. The fork-like
  structures prove the existence of a vortex-antivortex pair with charges
  $\pm 1$.
}
\label{fig4}
\end{figure}

In figure \ref{fig4}, we expose the phase structure 
of the wavefunction of the example at a particular propagation distance $z$.
The plots prove that the two dark spots of figure \ref{fig3} correspond indeed to a vortex-antivortex
pair.

Concerning the reconversion into a bright soliton \cite{paredes2014coherent}, 
we notice that it can  
take place 
when the excitation
reaches the boundary of the liquid of light as a single dark pulse. On the other hand,
when it does so as a vortex-antivortex pair, two waves propagating in opposite
directions along the
edge of the large soliton  get excited \cite{suppl}.

It must be emphasized that the generation of vortex and antivortex is
only one of the possible qualitative outcomes that emerge depending on
the relative velocity and phase. As in \cite{paredes2014coherent}, 
the droplets can simply coalesce into one. The collision can also result
in rarefaction pulses of different energies and speeds.
For low velocities, part of the energy can bounce back evolving into
a smaller bright soliton.
In all cases, surface and bulk sound waves are excited during the process.
If the collision is very violent, the large
soliton can be severely distorted, ceasing to be a liquid-like approximately homogeneous medium.

We close this section by noting that there are vortex solutions of the cubic-quintic 
equation (\ref{CQeq}) of the form
 $\psi=e^{i\beta z}e^{il\theta}f(r)$ with $\lim_{r\to\infty}f(r)=0$, where
$l$ is the topological charge and $\theta$ is the polar angle.
Their profiles and stability have been studied in 
\cite{quiroga1997stable,towers2001stability,berezhiani2001dynamics,
malomed2002stability,michinel2004square}
and their collisional dynamics in \cite{paz2004collisional}. 
We remark that the vortices that we are studying in this paper
as solutions of Eq. (\ref{EQtravel}) are different objects:
they live within the vorticity-less liquid of light and they only exist in pairs and
moving with a finite velocity.

\section{Collisions of vortex-antivortex pairs}

\begin{figure}[h!]
\begin{center}
\includegraphics[width=\columnwidth]{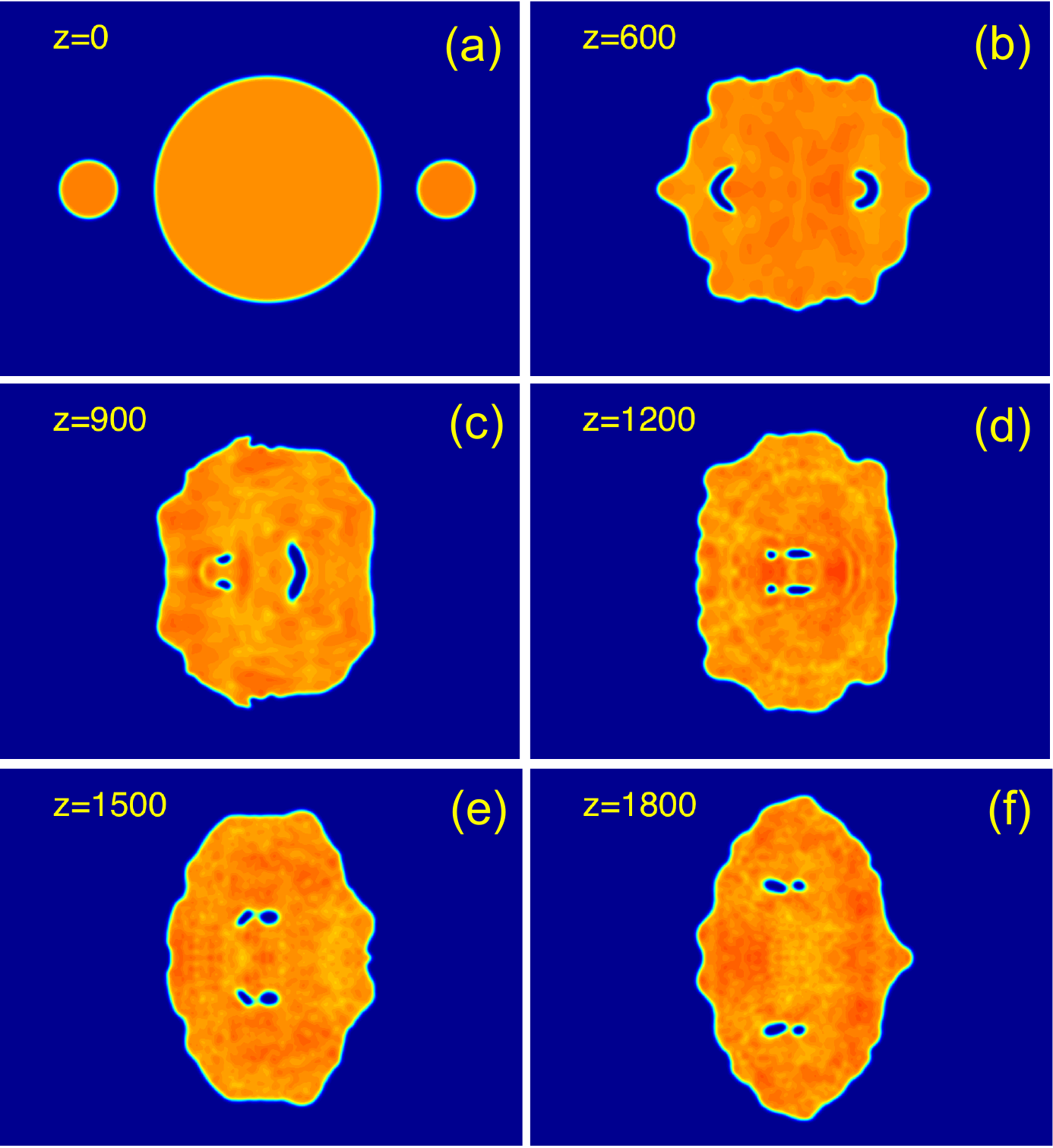}
\end{center}
\caption{Head-on encounter of two vortex-antivortex pairs resulting in an exchange mode.
Initial conditions are defined by Eq. \ref{init} with
$-{\bf x_2}={\bf x_3}=(180,0)$, ${\bf v_2}=-{\bf v_3}=(0.2,0)$, $\phi_2=5.3$, $\phi_3=5$.
The large soliton is the one with $\beta_1=0.1856$ and the smaller ones have $\beta_2=\beta_3=0.1802$.
Color code and axes are defined as in Fig. \ref{fig3}.
}
\label{fig5}
\end{figure}

Let us start illustrating the interactions 
by computing the head-on encounter of two vortex pairs created as 
described in section III. A typical example is displayed in Fig. \ref{fig5}. 
The result is an exchange in which the vortex of each pair recombines with the
antivortex of the other one (the exchange of a single vortex with a vortex-antivortex
pair was described in \cite{smirnov2015scattering} with cubic nonlinear potential.)
The solitary waves come out perpendicular to
the incoming direction. 
This can be understood in terms of the flow lines of Eq. (\ref{flow}), considering that, during the
approach, each pair generates a smooth inhomogeneity in the background in which the other
one propagates \cite{mironov2013scattering}. 
For instance, the antivortex on the top right (see the panels (c) and (d) of Fig. \ref{fig5}) 
feels the flow lines generated by the phase structure of the
vortex on the top left (see Fig. \ref{fig4}) 
and is pushed upwards. Conversely, the vortex in the
bottom right turns downwards because of the antivortex in the bottom left. Since these bends
tend to associate again vortex and antivortex, the propagation can continue after the exchange.
In Fig. \ref{fig5}, we have considered slightly different phases for the initial solitons in order to 
show that a perfect symmetry is not needed for this process.

\begin{figure}[h!]
\begin{center}
\includegraphics[width=\columnwidth]{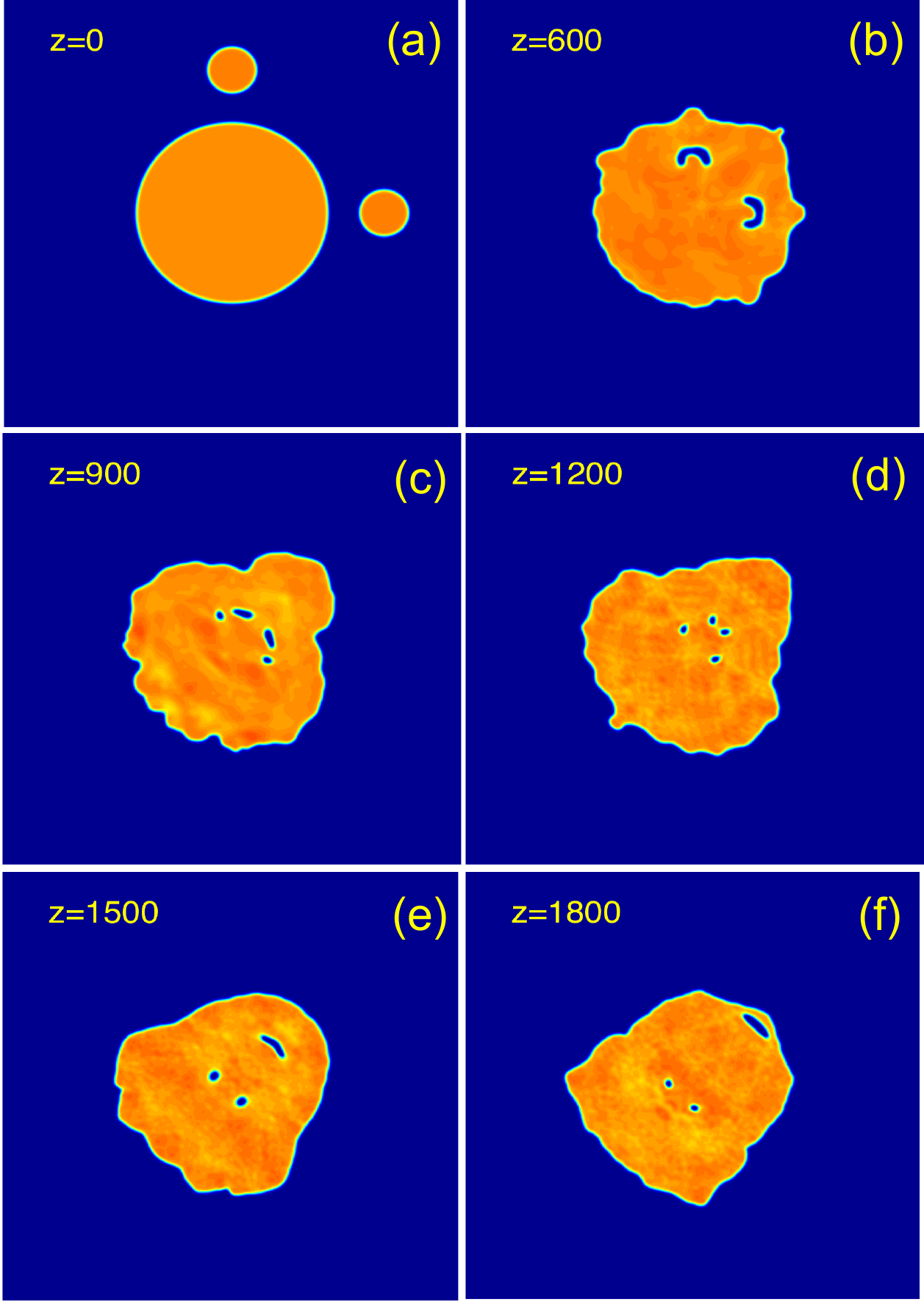}
\end{center}
\caption{Encounter at a right angle of two vortex-antivortex pairs resulting in an exchange mode.
Initial conditions are defined by Eq. \ref{init} with
${\bf x_2}=(0,180)$, ${\bf x_3}=(180,0)$, ${\bf v_2}=(0,-0.2)$, ${\bf v_3}=(-0.2,0)$, $\phi_2=\phi_3=5$.
The large soliton is the one with $\beta_1=0.1856$ and the smaller ones have $\beta_2=\beta_3=0.1802$.
The color code is defined as in Fig. \ref{fig2}.
The range of the axes is $x,y \in [-270,270]$.
}
\label{fig6}
\end{figure}

Similar exchanges can happen for collisions at angles. Figure \ref{fig6} depicts an example
where the incoming excitations are perpendicular to each other.
In this case, the vortex  moving downwards and the antivortex 
moving leftwards attract each other and coalesce into a dark blob which can be considered
an excited version of a rarefaction pulse. It comes out heading  the top right of the plot and
is finally reconverted into a (highly excited) bright soliton when it reaches the edge of the
medium. 
The remaining vortex and antivortex eventually couple to each other and continue to propagate
towards the bottom left. Notice that the velocity of this pair is much lower than that of the
aforementioned rarefaction pulse, as expected from the stationary solutions 
characterized in section II.

The simulation of Fig. \ref{fig6} is also interesting because it shows other generic features
of the dynamics, which can be better appreciated in the animation presented in the
supplemental material \cite{suppl}. In particular, we must emphasize that the evolution of the 
dark excitations is not elastic, in the sense that some energy is radiated away in the form
of sound waves. Moreover, faint rarefaction pulses, of small energy regarding Eq. (\ref{pE2}),
can be generated. These radiation processes take place during collisions  and also during the
relaxation of the coherently generated dark bubbles towards their stationary 
vortex pair form.

We also remark that, for encounters like that of Fig. \ref{fig6}, small changes in the initial
conditions can determine how the dark regions combine and greatly affect the outcoming
pulses. For instance, if we just change ${\bf v_3}$ from (-0.2,0) to (-0.21,0), therefore breaking
the symmetry, between both incoming vortex pairs,
the one moving horizontally arrives first. Instead of performing a exchange with the other vortex, 
it merges with the antivortex, creating an elongated void of net vorticity -1. This snake-like structure
starts rotating and eventually decays emitting a rarefaction pulse. We present this evolution
in \cite{suppl}. Thus, the encounter gives rise to a vortex-antivortex pair and a rarefaction pulse, just
as in Fig. \ref{fig6}, but their resulting propagation directions are rather different.
This simulation also shows that, when there is an eventual dark-bright reconversion, the outgoing
dark soliton does not necessarily come out with the same propagation direction as the dark
blob which generates it.

\begin{figure}[h!]
\begin{center}
\includegraphics[width=\columnwidth]{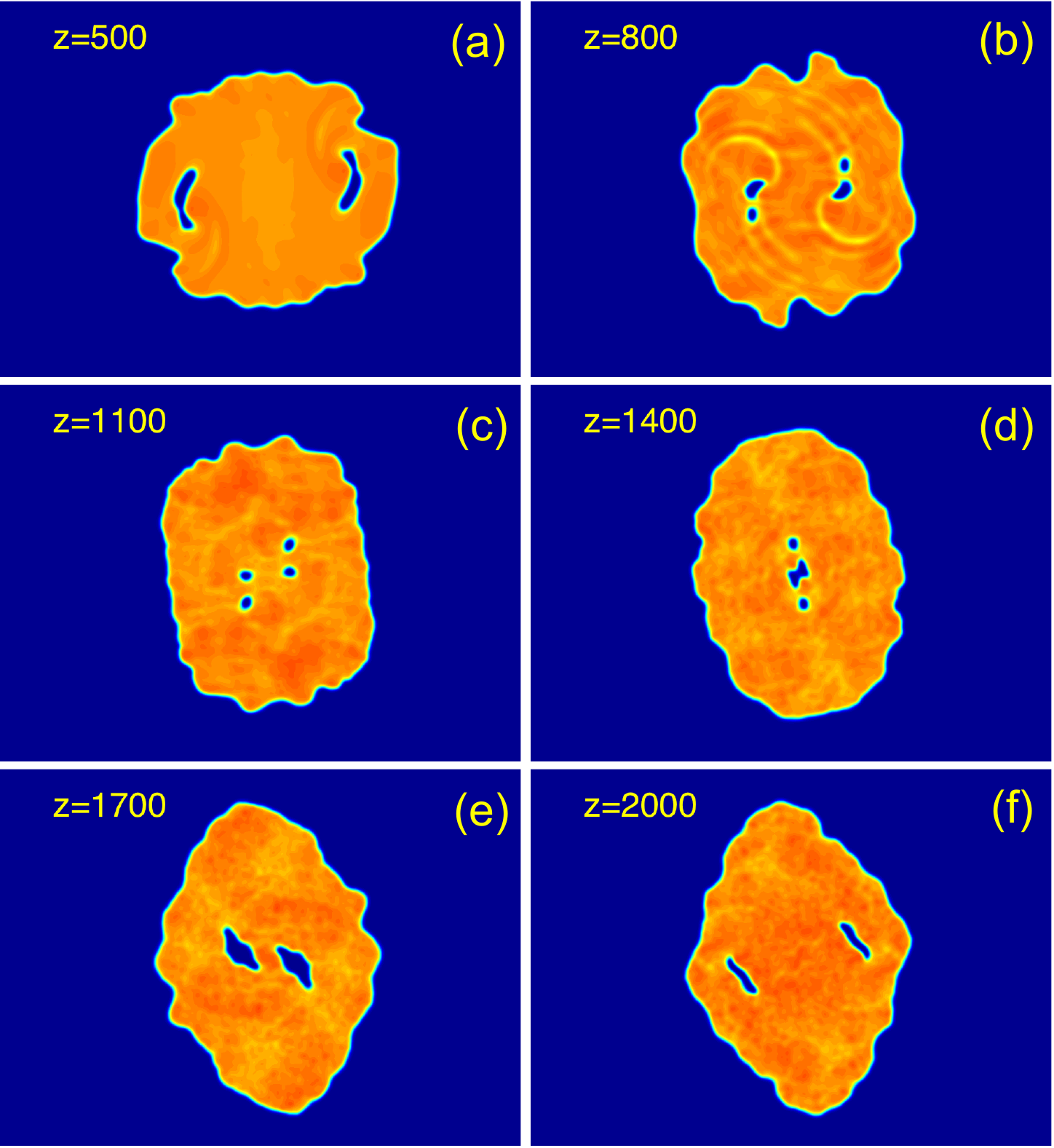}
\end{center}
\caption{Encounter of two vortex-antivortex pairs, shifted with respect to each
other along the direction transverse to propagation, resulting in pseudo-elastic
scattering.
Initial conditions are defined by Eq. (\ref{init}) with
${\bf x_2}=(-180,-10)$, ${\bf x_3}=(180,10)$, ${\bf v_2}=(0.2,0)$, ${\bf v_3}=(-0.2,0)$, $\phi_2=\phi_3=5$.
The large soliton is the one with $\beta_1=0.1856$ and the smaller ones have $\beta_2=\beta_3=0.1802$.
Color code and axes range are defined as in Fig. \ref{fig3}.
}
\label{fig7}
\end{figure}

We close this section by considering a head-on encounter in which the vortices of each pair 
meet each other (instead of heading an antivortex as in Fig. \ref{fig5}). This can be accomplished
by slightly shifting the $y$-position of the bright solitons defined in the initial conditions. 
An example is depicted in Fig. \ref{fig7}.

This evolution can be qualitatively understood noting that
 the vortices repel each other and therefore are slowed down while the
antivortices continue advancing. This induces a rotation of the whole dark structure, which eventually
breaks down resulting in two separate pulses which come out at an angle,
different from the incoming one. 
This is a kind of pseudo-elastic collision.
Notice, however, that the scattered pulses cannot be neatly considered vortex-antivortex as the
incoming ones. Vortex pairs and rarefaction pulses can be cleanly defined for stationary situations but
in dynamical evolutions like the present one, the separation between both is not obvious
and they can even transform into each other, as noticed in \cite{smirnov2015scattering} 
in a different  but somewhat related scenario.

\section{Collision of a rarefaction pulse with a vortex-antivortex pair}

\begin{figure}[h!]
\begin{center}
\includegraphics[width=\columnwidth]{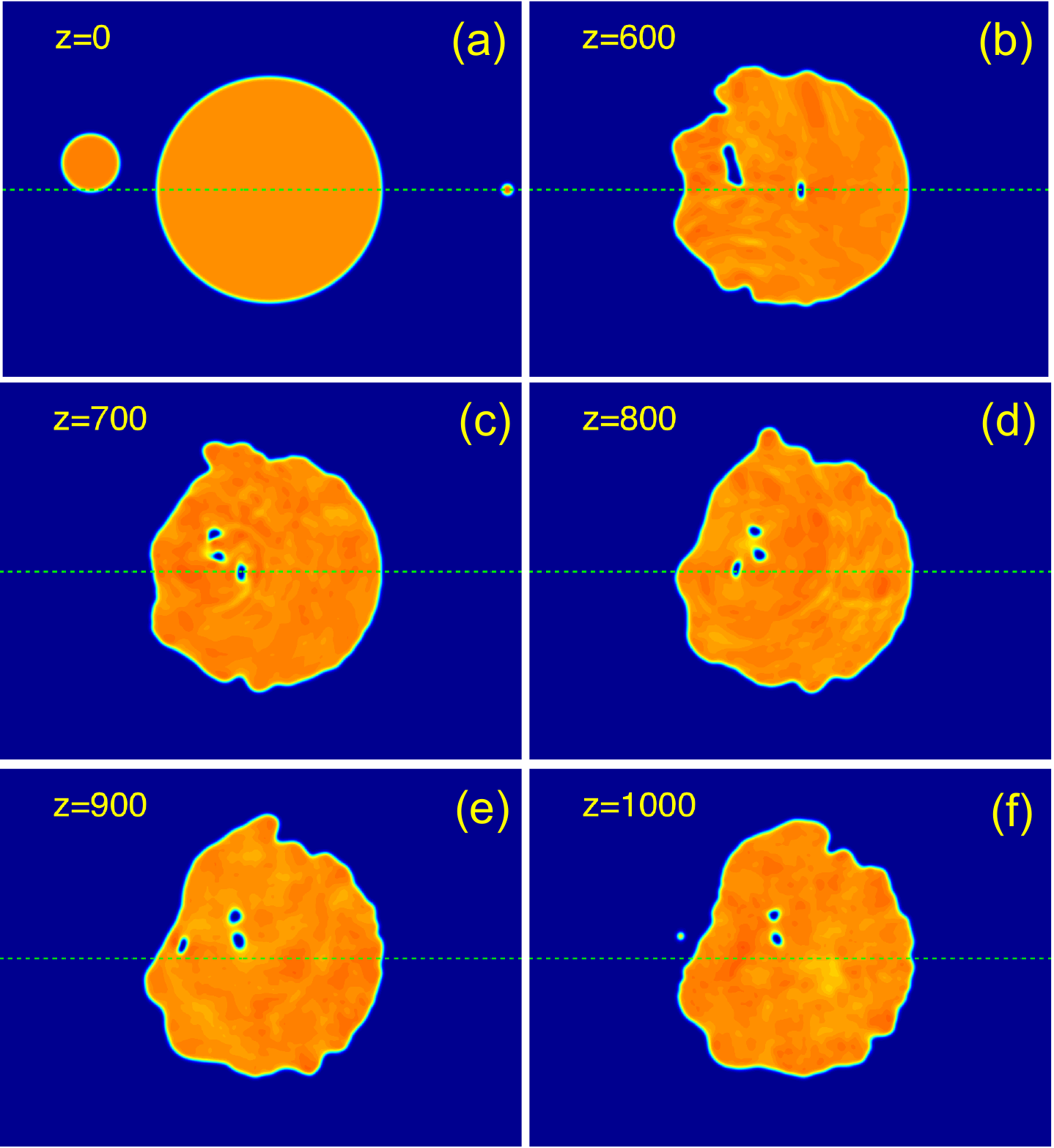}
\end{center}
\caption{(Color online) Flyby encounter of a rarefaction pulse with a vortex-antivortex pair.
The pulse trajectory is modified because of the flows generated by the vortex-antivortex phase
structure. The horizontal dashed line marks $y=0$, the path that the rarefaction pulse would
follow in the absence of other excitations.
Initial conditions are defined by Eq. (\ref{init}) with
${\bf x_2}=(-180,27)$, ${\bf x_3}=(240,0)$, ${\bf v_2}=(0.2,0)$, ${\bf v_3}=(-0.5,0)$, $\phi_2=4.9$,
$\phi_3=4$.
The large soliton is the one with $\beta_1=0.1856$ and the smaller ones have $\beta_2=0.1802$
and $\beta_3=0.15$.
Color code and axes range are defined as in Fig. \ref{fig3}.
}
\label{fig8}
\end{figure}

We now consider the encounter of a rarefaction pulse with a vortex-antivortex pair.
An illustrative case is sketched in Fig. \ref{fig8}. In the example, the dark regions moving in opposite
directions pass near each other but do not experience a direct contact. They keep their distinct
identities during the whole evolution and therefore this process is very similar to a elastic scattering.
The pulses continue their propagation away from each other and therefore we call this a flyby 
mode, following \cite{smirnov2015scattering}.
In the figure, it can be appreciated that the propagation of the rarefaction pulse is rotated by a small
angle when both waves meet (the horizontal dashed line has been included in the plots to guide the
eye). Again, this is due to the flow lines defined in Eq. (\ref{flow}) and
represented in Fig. \ref{fig2}, whose structure explains why the caviton turns upwards.
The vortex-antivortex pair is also affected by the encounter, by since its energy and momentum
(\ref{pE2}) is quite larger that that of the rarefaction pulse, it is much harder to appreciate the
diversion. Notice that this flyby mode is only relevant for a narrow window of 
the scattering impact parameter. If the caviton pulse moves far from the dipolar structure, the
phase gradients are tiny and their effect is negligible. On the other hand, if both waves are too near,
the dark regions recombine giving rise to more complicated evolutions, as we show in the next example.

\begin{figure}[h!]
\begin{center}
\includegraphics[width=\columnwidth]{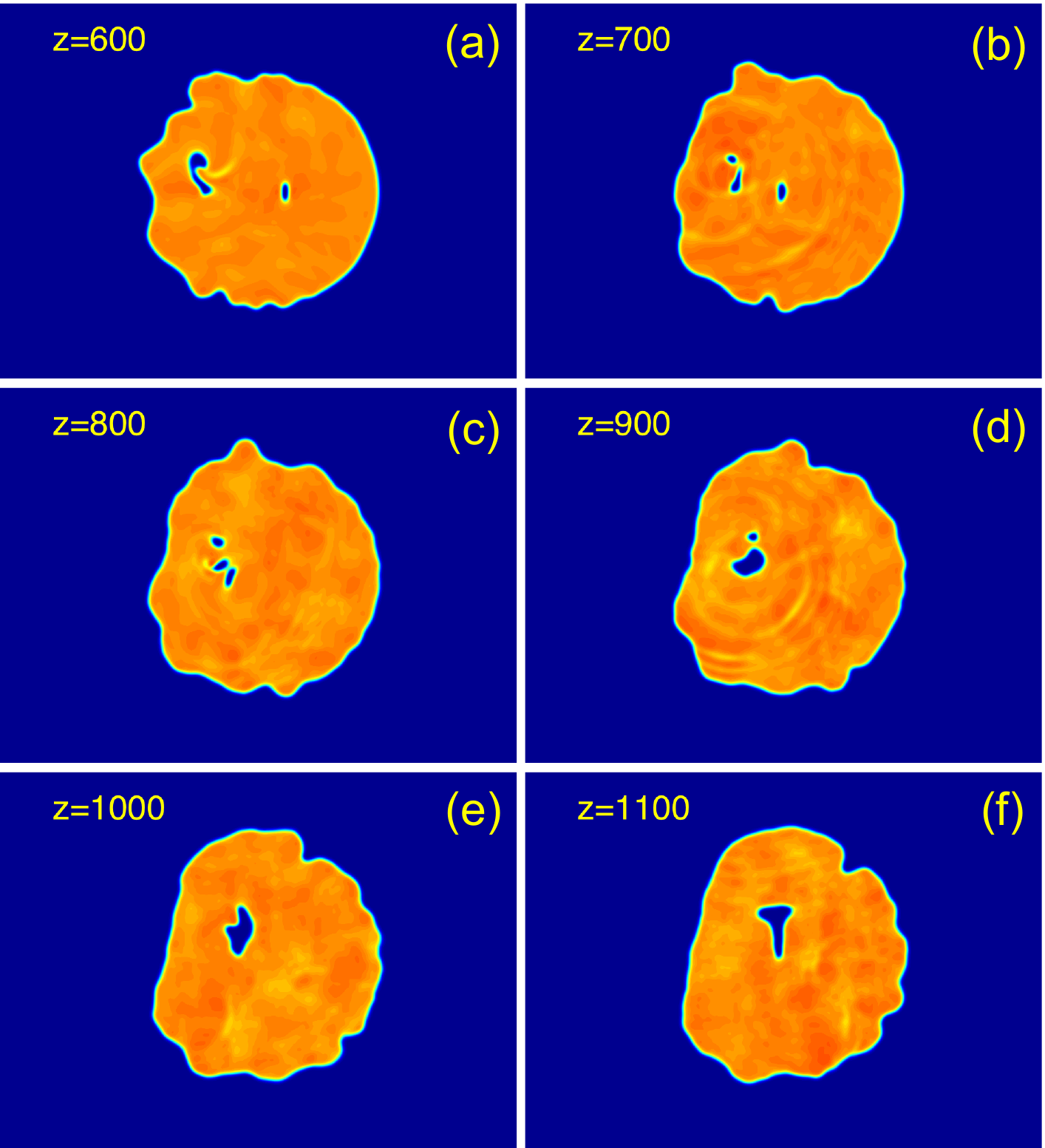}
\end{center}
\caption{(Color online) 
Inelastic collision of a rarefaction pulse and a vortex-antivortex pair.
The rarefaction pulse touches the antivortex, blends with it and, eventually, also gets
connected to the dark region around the vortex. The resulting  structure can
be considered as a highly excited vortex-antivortex pair which continues to propagate within
the liquid of light.
Initial conditions are defined by Eq. (\ref{init}) with
${\bf x_2}=(-180,18)$, ${\bf x_3}=(240,0)$, ${\bf v_2}=(0.2,0)$, ${\bf v_3}=(-0.5,0)$, $\phi_2=4.9$,
$\phi_3=5$.
The large soliton is the one with $\beta_1=0.1856$ and the smaller ones have $\beta_2=0.1802$
and $\beta_3=0.15$.
Color code and axes range are defined as in Fig. \ref{fig3}.
}
\label{fig9}
\end{figure}

The initial conditons in Fig. \ref{fig9} resemble those of Fig. \ref{fig8}, but the 
initial $y$-displacement of the bright solitons is slightly smaller, yielding
a smaller impact parameter for the collision of the dark waves.
In this case, the dark regions associated to the antivortex and the rarefaction pulse
come into contact and merge, initially giving rise to a large blob of vorticity -1. 
Since the vortex-antivortex pair has the larger momentum
and energy, the subsequent evolution can be roughly described as an absorption of the 
rarefaction pulse by the pair, which becomes highly excited, but continues its propagation rigthwards. 
This structure slowly relaxes towards the stationary vortex-antivortex solution by the emission of
sound waves and faint rarefaction pulses \cite{suppl}.
In \cite{suppl}, we also present a simulation in which the vortex pair and the caviton approach each
other with zero impact parameter. Roughly, the dynamics can be understood in terms of the 
previous discussion: when the dark regions touch each other, the rarefaction pulse is swallowed by 
the vortex-antivortex which, albeit excited, continues its propagation.
We have checked that this kind of
qualitative
 behavior is quite generic, regardless of the incoming angles and velocities.

\section{Collisions of rarefaction pulses}

Let us now discuss the case of two interacting rarefaction pulses.
First of all, we notice the existence of flyby modes, similar to those described in the previous section,
when the impact parameter is not too large but enough to avoid direct contact.

\begin{figure}[h!]
\begin{center}
\includegraphics[width=\columnwidth]{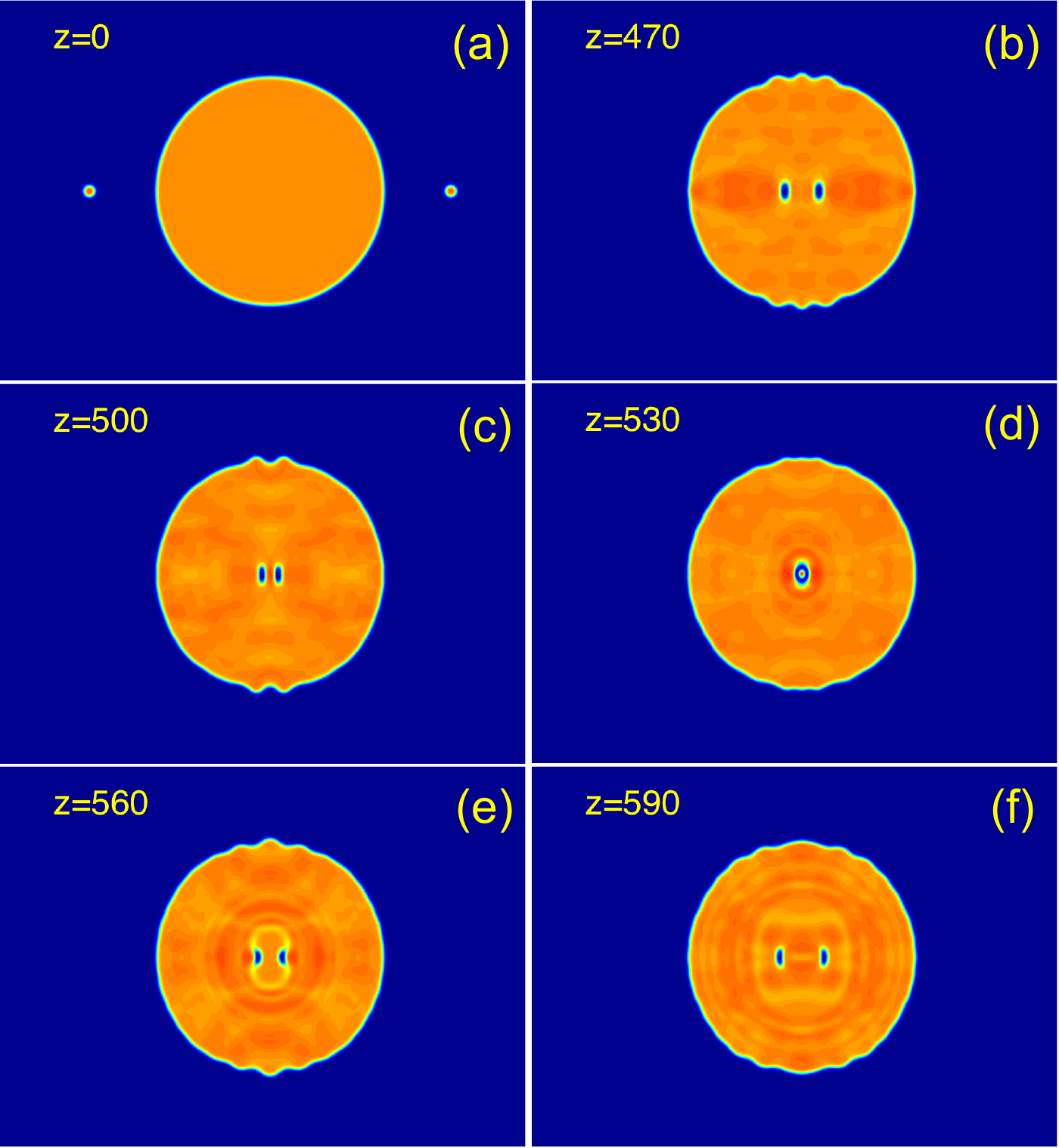}
\end{center}
\caption{(Color online) 
Symmetric head-on encounter of two rarefaction pulses which cross each other, losing a fraction of their
energy in the process.
Initial conditions are defined by Eq. (\ref{init}) with
$-{\bf x_2}={\bf x_3}=(180,0)$, ${\bf v_2}=-{\bf v_3}=(0.5,0)$,  $\phi_2=\phi_3=6$.
The large soliton is the one with $\beta_1=0.1856$ and the smaller ones have  $\beta_2=\beta_3=0.15$.
Color code and axes range are defined as in Fig. \ref{fig3}.
}
\label{fig10}
\end{figure}

It is also worth commenting on the
 dynamics of head-on collisions. 
The most common result is illustrated in Fig. \ref{fig10}. When the pulses meet, a larger dark blob is
created with, possibly, a bright spot inside (see panel (d) of Fig. \ref{fig10}.) 
Then, two rarefaction pulses appear again and continue their propagation. During the encounter,
part of the energy is radiated away and, therefore, the pulses after the collision are slightly
fainter and faster. Thus, in this respect, the rarefaction pulses behave as dark quasi-solitons.
We remark that this happens for symmetric encounters as the one of the figure or asymmetric
ones with pulses of different energies.
As expected, when the cavitons reach the edge of the large soliton, they can be reconverted in 
bright solitons again. In fact, the simulation of Fig. \ref{fig10} can be interpreted as a
bright-dark-bright-dark-bright transformation of the propagating excitation \cite{suppl}.

\begin{figure}[h!]
\begin{center}
\includegraphics[width=\columnwidth]{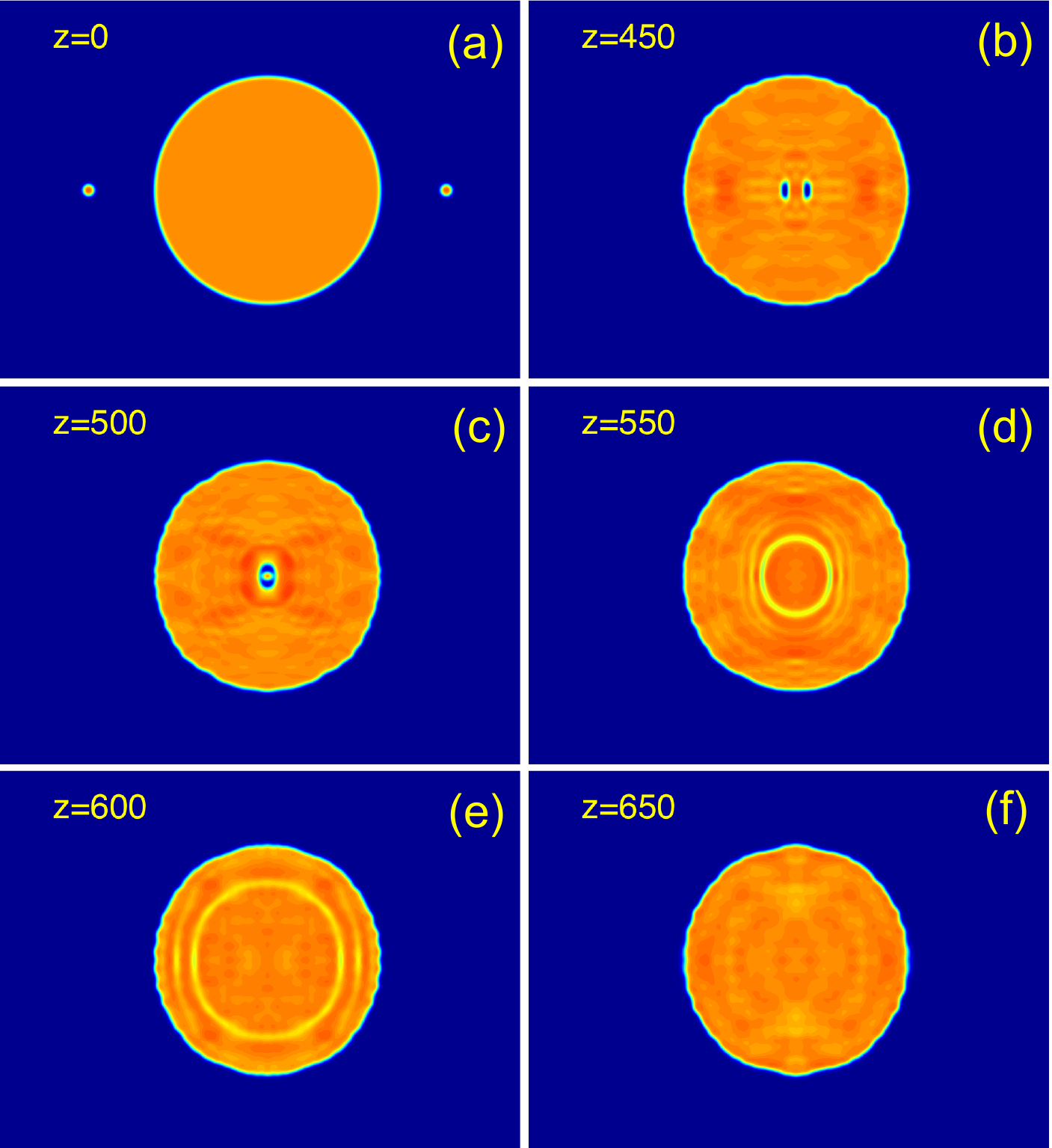}
\end{center}
\caption{(Color online) 
Symmetric head-on encounter of two rarefaction pulses resulting
in a fully inelastic collision. The pulses annihilate each other and yield all their energy to a 
circular sound wave.
Initial conditions are defined by Eq. (\ref{init}) with
$-{\bf x_2}={\bf x_3}=(180,0)$, ${\bf v_2}=-{\bf v_3}=(0.5,0)$,  $\phi_2=\phi_3=4.8$.
The large soliton is the one with $\beta_1=0.1856$ and the smaller ones have  $\beta_2=\beta_3=0.15$.
Color code and axes range are defined as in Fig. \ref{fig3}.
}
\label{fig11}
\end{figure}

Curiously, the picture changes completely if the initial conditions are properly fine tuned.
Figure \ref{fig11} depicts an example in which the rarefaction pulses annihilate each other and
their energy is radiated in the form of a circular sound wave.
Visibly, the behavior of the rarefaction pulses in this case totally differs from that of form
preserving solitons. As a matter of fact, the seemingly antagonistic character of Figs.
\ref{fig10} and \ref{fig11} can be continuously connected, by noticing
that in all head-on encounters the outgoing energy is shared by a bulk wave and two rarefaction pulses.
In Fig. \ref{fig10}, most of the energy goes to the latter whereas in Fig. \ref{fig11} it is mostly acquired by
the former, while other initial conditions lead to intermediate possibilities.

\begin{figure}[h!]
\begin{center}
\includegraphics[width=\columnwidth]{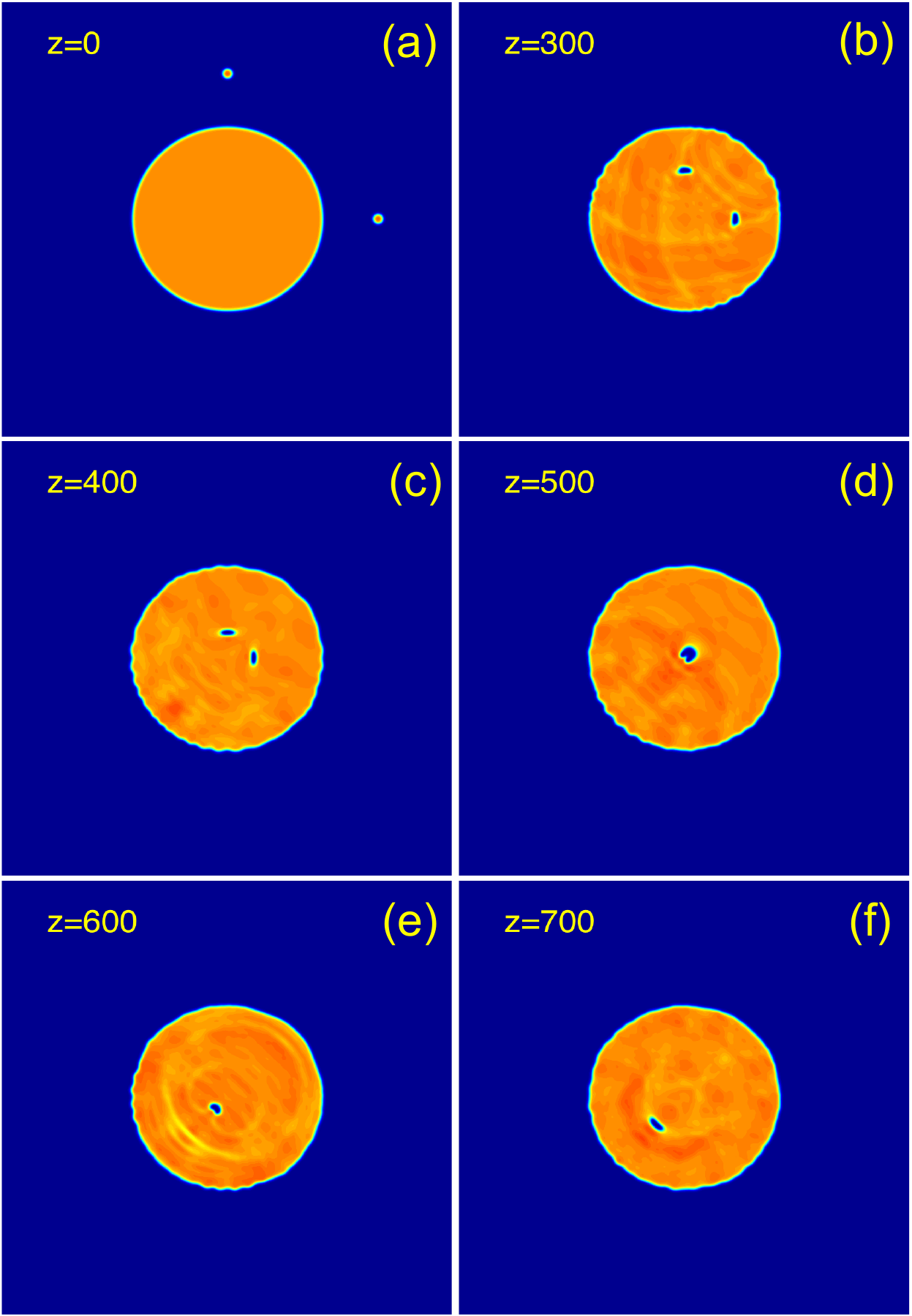}
\end{center}
\caption{(Color online) 
Two rarefaction pulses collide perpendicularly and merge.
Initial conditions are defined by Eq. (\ref{init}) with
${\bf x_2}=(0,180),{\bf x_3}=(180,0)$, ${\bf v_2}=(0,-0.5)$, $ {\bf v_3}=(-0.5,0)$,  $\phi_2=\phi_3=5$.
The large soliton is the one with $\beta_1=0.1856$ and the smaller ones have  $\beta_2=\beta_3=0.15$.
The color code is defined as in Fig. \ref{fig2}.
The range of the axes is $x,y \in [-270,270]$.
}
\label{fig12}
\end{figure}

Finally, we comment on the encounter of rarefaction pulses at angles. Figure \ref{fig12} 
illustrates this case by considering a perpendicular concurrence.
As in the previous cases, the dark regions combine producing a dark blob, which is larger than the 
incoming ones. However, in this case this blob can survive and, in a loose sense, propagate in the direction
required by momentum conservation. 
Thus, the simulation of Fig. \ref{fig12} can be neatly portrayed as the merging of two rarefaction pulses
into a more energetic one.
Similarly to all of the presented examples, part of the energy is radiated away during the process.

We close the section by noticing that there is a second typical qualitative behavior, which we show
in the last animation of \cite{suppl}. What happens there is that the dark blob splits  giving
rise again to two rarefaction pulses (we emphasize that, even if in \cite{suppl} it may seem that
the dark pulses coming out of the collision
propagate almost in parallel, they are not vortex and antivortex).
Roughly, this last possibility can be thought of as another example of quasi-elastic scattering or
as a bounce of the pulses against each other.

\section{Summary and outlook}

In this work, we have numerically analyzed Eq. (\ref{CQeq}), reporting on a number of novel
qualitative phenomena for the cubic-quintic model in 1+2 dimensions.

The interplay of diffraction with focusing and defocusing nonlinear effects endows the cubic-quintic
nonlinear Schr\"odinger equation with an extremely rich phenomenology. In particular, there are 
dark traveling waves and
bright solitons, which for large powers become liquid-like. 
Noticeably, the dark and bright stationary and stable solitary waves can transform into each other
during evolution. In particular, a bright soliton can excite a rarefaction pulse when it meets
a bright soliton of larger power \cite{paredes2014coherent}. We have shown that a vortex-antivortex
pair can be generated in a similar way. The process, however, is not as clean as in the 
previous case. The incoming soliton has to be larger and gives rise to a more pronounced distortion
of the flat-top soliton. Moreover, the vortex and antivortex are not generated directly, but only 
as the end result of a snake instability of an initial dark blob. Thus, the radiation of part of the excess
energy is essential in approaching the stationary vortex-antivortex solution.
When a strong enough rarefaction pulse reaches the border of the liquid of light, it typically generates an outgoing
bright soliton. On the other hand, the vortex-antivortex pair excites a
couple of
surface waves propagating in opposite directions.

The possibility of creating the dark states by interference and nonlinear evolution has allowed us to 
propose numerical experiments concerning their scattering with initial conditions which only
include bright solitons, see Eq. (\ref{init}). 
We have made a qualitative analysis of the encounters between vortex-antivortex pairs and rarefaction
pulses. In brief, our results can be summarized as follows:

\begin{itemize}

\item If the vortex of a pair meets an antivortex of another pair and viceversa, they tend to get
exchanged resulting in two new pairs with different propagation directions.

\item If the impact parameter of a collision is large enough and the dark regions do not touch each
other, there are elastic flyby modes and the propagation direction of each wave is altered because of the flow
lines associated to the opposite structure.

\item When a vortex or a rarefaction pulse touches a vortex-antivortex pair, an excited dark blob is created.
It propagates for a while and eventually decays approaching the stationary states. The end result is strongly
dependent on initial conditions.

\item Rarefaction pulses which collide head-on typically cross each other, losing some energy by 
radiating sound waves. In particular situations, the radiation can take most of the energy. If the
pulses collide at an angle, they can merge into a larger rarefaction pulse or scatter quasielastically.

\end{itemize}

This list does not exhaust the possibilities but it certainly provides a qualitative description for most
of the collisions between dark traveling waves. 
It is tempting to interpret the traveling waves as quasiparticles and to try to understand collisions
in terms of their energy-momentum conservation, Eq. (\ref{pE2}).
Implicitly, this has been our point of view when using the words ``elastic'' and
``inelastic''.
Notice that $p$ and $E$ as a whole are conserved in a collision. Nevertheless, if we only take into account
the dark traveling waves, the conservation breaks down, as it obvious from figure \ref{fig10}.
The main reason is that sound waves take a sizable fraction of energy and momentum in many
processes. Moreover, as we have already emphasized, the dark waves typically appear in
excited form and therefore the velocity-momentum and 
dispersion relations that can be deduced from the stationary 
solutions only apply approximately. Excited dark states have complicated
dynamics and cannot always be easily identified with their stationary counterparts.
Thus, the quasiparticle interpretation is illustrative but it should be clear that 
it is just a qualitative
rough description.

Our results open some interesting possibilities. First of all, it would be nice to realize the 
described phenomena in  optical setups along the lines of
 \cite{falcao2013robust,wu2013cubic,wu2015solitons}. It would also be desirable to study similar effects
 in two dimensions for the cubic defocusing nonlinearity, since it is relevant for Bose-Einstein experiments
 like \cite{proud2016jones}, see also \cite{verma2016snake} and references therein. 
 Moreover,
 it would be worth considering the three dimensional cubic-quintic case,
 which  supports top-flat stable spatiotemporal solitons
 \cite{desyatnikov2000three,jovanoski2001light} and vortices 
 \cite{desyatnikov2000three,mihalache2002stable}.
Their collsional dynamics has been analyzed in 
\cite{hong2008energy,adhikari2016elastic}
but the dynamics of dark traveling waves has not been described yet.
Using the cubic defocusing Schr\"odinger equation,
interesting dynamical
 analysis of the interplay of rarefaction pulses, vortex rings and vortex lines in 1+3 dimensions
have been presented in the context of Bose-Einstein condensates 
\cite{berloff2002evolution,berloff2004interactions,komineas2005collisions}
and superfluids 
\cite{caplan2014scattering}. It would be desirable to make contact with these analyses 
in the cubic-quintic case.
Finally, we remark that our setup has partial
 similarities with other physical systems as, {\it e.g.}, the scattering by impurities in superfluids as
 recently modeled in \cite{pshenichnyuk2016inelastic}. It could be worth exploring analogies between 
 different frameworks.

\acknowledgments

We thank David N\'ovoa and Jos\'e Ram\'on Salgueiro for
useful comments. 
This work is supported by grants FIS2014-58117-P
from Ministerio de Econom\'\i a y Competitividad and grants GPC2015/019
and EM2013/002 from Xunta de Galicia.
The work of D. F. is supported by the FPU Ph.D. program

\end{document}